\documentclass[twocolumn,showpacs,preprintnumbers,amsmath,amssymb,prl]{revtex4}
\usepackage{graphicx}
\usepackage{dcolumn}
\usepackage{bm}
\usepackage{color}
\begin{document}

\title{Model independence in two dimensions and polarized cold dipolar
  molecules}

\author{A.G. Volosniev, D.V. Fedorov, A.S.~Jensen, and N.T. Zinner}

\affiliation{ Department of Physics and Astronomy,
         Aarhus University, DK-8000 Aarhus C, Denmark }

\date{\today}

\begin{abstract}
We calculate the energy and wave functions of two
particles confined to two spatial dimensions interacting via arbitrary
anisotropic potentials with negative or zero net volume.  The general
rigorous analytic expressions are given in the weak coupling limit where
universality or model independence are approached.
The monopole part of anisotropic potentials is crucial in
the universal limit.  We illustrate the universality 
with a system of two arbitrarily
polarized cold dipolar molecules in a bilayer.  We discuss the
transition to universality as function of polarization and binding
energy, and compare analytic and numerical results obtained by the
stochastic variational method.  The universal limit is essentially
reached for experimentally accessible strengths.
\end{abstract}


\maketitle

\paragraph*{Introduction.} 
The existence of bound quantum states in a potential is a problem that appears
in many scientific disciplines. While in three dimensions it is known
that sufficiently shallow potentials will not bind, for one- and
two-dimensional non-positive potentials early arguments by Landau and
Lifshitz \cite{landau} demonstrate that at least one bound state is
always present for arbitrarily small potential strength.  Only much
later this was generalized to potentials of negative or zero spatial volume by
Simon \cite{simon1976}.  Such bound states and their structures are
not only of formal interest but serve a great number of applications
in several areas including exotic atoms \cite{exotic},
excitons in carbon nanotubes \cite{carbon}
or semiconducting microcavities \cite{micro} and organic interfaces
\cite{organic}, pairing in two-dimensional Fermi gases \cite{pairing},
localization of adatoms on surfaces \cite{adatoms} or
through nanosctructuring \cite{nanostructured}, and in population genetics
\cite{genetics}.

In this letter we introduce a general
formalism to calculate the leading, as well as higher, order terms for
arbitrary anisotropic weak potentials.  This opens the door for many
new applications since physical systems usually have anisotropic
features that make models assuming cylindrical symmetry inaccurate.
We test our formulas in the
emerging field of cold polar molecules \cite{review} where layered
geometries have recently been experimentally realized
\cite{layerexp}. The long-range dipolar potential can be made
anisotropic by tilting the external aligning field.  Novel
superfluid and density-wave instabilites have been proposed
in this geometry \cite{supercdw}.  
However, very little is known about few-body bound
states in the system, particularly in the experimentally relevant weak
coupling limit, where universal features emerge independent of
specific models, and previous works have only considered the case of 
perpendicular polarization \cite{arm10a,san10,baranov2010}.  
We shall compare analytic and numerical results and
investigate the approach towards the universal limit.  We deduce some
implications for three-body \cite{nie99} as well as for many-body
states in two dimensions (2D).

\paragraph*{Formulation.}
Consider two particles confined to two spatial dimensions and interacting via
a pair potential, $V(\vec r)$, where $\vec r$ is the relative
coordinate.  Using polar coordinates, $\vec r = (x,y) = (r\cos
\varphi, r\sin \varphi)$, we can write the Schr\"{o}dinger equation as
\begin{equation} \label{e30}
\left[-\frac{1}{s}\frac{\partial}{\partial
s}{s}\frac{\partial}{\partial
s}- \frac{1}{s^2}\frac{\partial^2}{\partial \varphi^2} +
 \lambda \frac{2\mu d^2}{\hbar^2} V(s,\varphi)\right]\psi=\alpha^2\psi \;,
\end{equation}
with $\psi$ the wave function, $\mu$ the reduced mass, $\lambda$
the dimensionless strength, $\alpha^2=2\mu d^2E/\hbar^2$, $E$ the
energy, $d$ the unit of length, and $s=r/d$ the reduced
coordinate.  The wave function can be partial-wave expanded,
\begin{equation} \label{e40}
\psi(s,\varphi)=\frac{1}{\sqrt{s}}\sum_{m=-\infty}^{m=\infty}
 a_m f_m(s)\exp(im\varphi) \;, 
\end{equation}
where $a_m$ are the expansion coefficients and the 2D radial
wave functions, $f_m(s)$, satisfy the system of equations
\begin{equation} \label{eq-w1} 
f_m'' +\frac{1-4m^2}{4s^2}f_m+\alpha^2f_m= \lambda\sum_{l}\frac{a_l}{a_m}f_lV_{ml}(s)\;.
 \end{equation}								
The matrix elements, $V_{ml}$,
\begin{equation} \label{eq-w2} 
V_{ml}=\frac{1}{2\pi} \frac{2\mu d^2}{\hbar^2}
 \int_0^{2\pi} e^{i(l-m)\varphi} V(s,\varphi)\mathrm{d}\varphi \;,
\end{equation}
carry all information about the potential.  For cylindrical
potentials $V_{ml} \propto \delta_{ml}$ and the different $m$-values
decouple in Eq.~\eqref{eq-w1}.

\paragraph*{General derivation.}
The regular radial solution to Eq.~\eqref{eq-w1} at the origin
provides the usual boundary condition for a centrifugal barrier
potential, i.e. we choose $\lim_{s\to0}s^{-1/2-|m|}f_m(\alpha,s)=1$,
where we inserted explicitly the dependence on the energy parameter,
$\alpha$, in $f_m$. 
We assume that the potential, and consequently also the right hand side of
Eq.~\eqref{eq-w1}, diverge slower than $1/s^2$ when $s\rightarrow 0$.
The integral form of the equations in
\eqref{eq-w1} is given as in \cite{new86},
\begin{equation}
\begin{split}&a_mf_m(\alpha,s)=a_mF_m(\alpha,s) \\&-
 \lambda \sum_{l}a_l\int_0^s g_{|m|}(\alpha,s,s')V_{ml}(s')f_l(\alpha,s')\,\mathrm{d}s'
\, , \label{eq-w4}
\end{split}
\end{equation}
where the boundary condition at $s=0$ is obeyed through the solution
of the free Schr\"{o}dinger equation:
\begin{equation} \label{eq-w5} 
F_m(\alpha,s)=\sqrt{s}J_{|m|}(\alpha s)(2/\alpha)^{|m|} |m|!\;, 
\end{equation}
where $J_m$ is the Bessel function and where $g_{|m|}$ in
Eq.~\eqref{eq-w4} is the Green's function given as
\begin{equation} 
\begin{split} g_{|m|}(\alpha,s,s')=\frac{i\pi}{4}\sqrt{ss'}
[H_{|m|}^{(1)}(\alpha s)H_{|m|}^{(2)}(\alpha s') \\ \label{eq-w9}
 -H_{|m|}^{(1)}(\alpha s')H_{|m|}^{(2)}(\alpha s)] \;
\end{split}
\end{equation}
in terms of the Hankel functions, $H_m^{(n)}$. 
For bound states (where $\alpha=i|\alpha|$)
both the free wave functions in Eq.~\eqref{eq-w5} and the
Green's function in Eq.~\eqref{eq-w9} diverge at large distances, but
their combination in Eq.~\eqref{eq-w4} must vanish exponentially.  This
provides the quantization condition for the energy.  Mathematically
the condition is given as 
\begin{equation} \label{add-1} \sum c_{ml}a_l= -a_m, \end{equation}
\begin{eqnarray} 
\label{eq-w11}
&& c_{ml} =  \\ \nonumber 
&& \lambda \left(\frac{\alpha}{2}\right)^{|m|}\frac{i\pi}{2|m|!} 
 \int_0^\infty\!\sqrt{s'}H_{|m|}^{(1)}(\alpha s') 
V_{ml}(s')f_l(\alpha,s') \mathrm{d}s' \;.
\end{eqnarray} 
This homogeneous set of linear equations \eqref{add-1}
has non-trivial solutions only if
\begin{equation} \label{add-11} \det(c_{ml}+\delta_{ml})=0, 
\end{equation}
where $\delta_{ml}$ is Kronecker's delta.  

\paragraph* {Solution in the limit of weak binding.}
We solve Eq.~\eqref{eq-w4} with the boundary condition (\ref{add-1}) 
as an expansion in $\lambda$ 
\begin{equation} \label{add-2} f_m=\sum_{n=0}^\infty \lambda^n f_m^{(n)},\;    
a_m=a_0\sum_{n=1}^\infty \lambda^n a_m^{(n)}, 
\end{equation}
where $a_0$ is a normalization constant multiplying all $a_m$ for $m\neq 0$.
In the lowest order we find $a_m^{(1)}$  from Eq.~\eqref{add-1}
using $f_{0}^{(0)}(0,s)=\sqrt{s}$ for $s|\alpha|\ll1$. 
Then Eq.~\eqref{eq-w4} gives $f_m^{(0)}$ and $f_0^{(1)}$,
\begin{eqnarray} 
 & f_0^{(1)}(0,s) = - \sqrt{s} \int_0^{s} s'V_{00}(s') \ln (s'/s) \mathrm{d} s' \;, \label{add-3}\\\label{add-4}
& f_m^{(0)}(0,s) = s^{|m|}\sqrt{s} +
\frac{\int_0^{s} \frac{(s')^{2|m|} - (s)^{2|m|}}{(ss')^{|m|}}
 s' V_{m0}(s') \mathrm{d} s'}{\int_0^{\infty} V_{m0}(s') (s')^{1-|m|}. \mathrm{d} s'}.
\end{eqnarray}
Solving Eq.~\eqref{add-11} in the leading order in $\lambda$ gives the energy
\begin{eqnarray}
& E= - \frac{2\hbar^2}{\mu d^2} \exp(-2\gamma -2/(\lambda^2A_0))
 \;, \label{e127}  \\    \label{e128} &  A_0 \equiv
 -\int_0^{\infty} \sqrt{s} V_{00}(s) f_0^{(1)}(0,s) \rm{d} s + \\ \nonumber
 & \sum_{m \neq 0}  \int_0^{\infty} \frac{s^{1-|m|}}{2|m|} V_{m0}(s)  \mathrm{d} s
 \int_0^{\infty} \sqrt{s'} V_{0m}(s') f_m^{(0)}(0,s') \mathrm{d} s' ,
\end{eqnarray}
where $\gamma$ is the Euler's constant.  Next order corrections
can be determined iteratively using the calculated $a_m^{(1)},
f_m^{(0)},f_0^{(1)}$ as described in \cite{vol11}.  The asymptotic
form of the radial wavefunctions for 2D and finite-range anisotropic 
potentials is given generally as
\begin{equation} \label{eq-w6}
f_{m}(\alpha,s) \rightarrow\sqrt{s}H_{|m|}^{(1)}(\alpha s) \delta_{m0}\;, 
|\alpha|s \ll 1. \;
\end{equation}
There is
always a bound state for very weak potentials even for zero net
volume, $\int r V \mathrm{d}r\mathrm{d}\varphi \leq 0$, and the
threshold behavior of the energy is given by Eq.~\eqref{e127}.  For
cylindrical potentials only the first term in Eq.~\eqref{e128}
contributes as $V_{0m}\propto \delta_{0m}$, but even the leading order
is still in general complicated.

\paragraph*{Dipole results.}
Our first application is the system of two polarized molecules of
reduced mass $\mu=M/2$ confined to two parallel planes separated by a
distance $d$. The corresponding dipole-dipole potential, $V$,
projected to this two-dimensional geometry is
\begin{equation} \label{e20}
 V(r,\varphi)=D^2 \frac{r^2+d^2-3(r\cos\varphi\cos\theta+
 d\sin\theta)^2}{(r^2+d^2)^{5/2}}\;,
\end{equation}
where $D$ is the dipole moment and $\theta$ is the polarization angle
measured from plane.  This potential is reflection invariant with zero
net volume for any polarization. It has monopole, dipole, and quadrupole
terms only.  The strength is now $\lambda \rightarrow U \equiv M
D^2/(d\hbar^2)$, the non-zero matrix elements are 
$V_{m,m}, V_{m,m\pm1}, V_{m,m\pm 2} $.  Contributions
to second order in $U$ on the right hand side of
Eq.~\eqref{e127} are included by using $m=0,\pm 1,\pm 2$ since higher
partial waves only contribute to the wave function through at least
third order in $U$.  We here systematically restrict ourselves
to first order, and arrive at the energy expression
\begin{equation} \label{e70}
  E  = - \frac{4 \hbar^2}{ M d^2} \exp\left(-2\gamma
 -\frac{2(1+U B_1)}{U^2(A_0 + U A_1)}\right) \; ,
\end{equation}
where the coefficients, $A_0,A_1$ and $B_1$, are defined by
\begin{eqnarray} \label{e80}
A_0 &=& \frac{1}{4}M_c^2 +\frac{1}{8}
\sin^2{2\theta}+\frac{1}{32}\cos^4{\theta} \;, \\ 
A_1 &=& + 0.0053  
\sin^2{2\theta}\cos^2{\theta} - 0.0033 \sin{2\theta}\cos^4{\theta}
 \nonumber \\  &-& 0.0019 \cos^6\theta
- M_c\big(0.0349 \sin^2(2\theta)   \\ \nonumber
&+& 0.0054 \cos^4(\theta) 
+ 0.0156 M_c \cos^2\theta + 0.0343 M_c^2\big) \;, \\ 
B_1 &=& -  1.204 M_c - \frac{1}{16} \cos^2\theta, \\ 
 M_c &=& \frac{3}{2}\sin^2(\theta)-\frac{1}{2}, ;\;
 \frac{a}{4d} = \exp\left(\frac{(1+U B_1)}{U^2(A_0 + U A_1)}\right).
\label{e90}
 \end{eqnarray}
Here $a$ is the scattering length which is a function of 
the strength and the polarization
angle \cite{nie01}.  The energy very close to threshold is exponential
in $U^2$, as seen in Eq.~\eqref{e127}, and determined by the
polarization angle through $A_0$.  The first order terms, $(A_1,B_1)$,
in $U$ exhibit the difference in approach to threshold for the
different signs of the strength, $U$.  The second order terms,
$(A_2,B_2)$, are necessary to get the correct $U$-independent
pre-exponential factor in the energy.  These higher order terms are
sums of a large number of contributions expressed as definite
integrals.  
The energy for $\theta=\pi/2$ has been considered in recent studies \cite{arm10a,san10,baranov2010}.
While our result is in agreement with Ref. \cite{baranov2010}
to second order, the approximation in Ref. \cite{san10} deviates
from our results in Eq.~\eqref{e70} in first order
through the $A_1$-term. We give more details on the
generalization to arbitrary polarization directions in \cite{vol11}.

\begin{figure}
\input{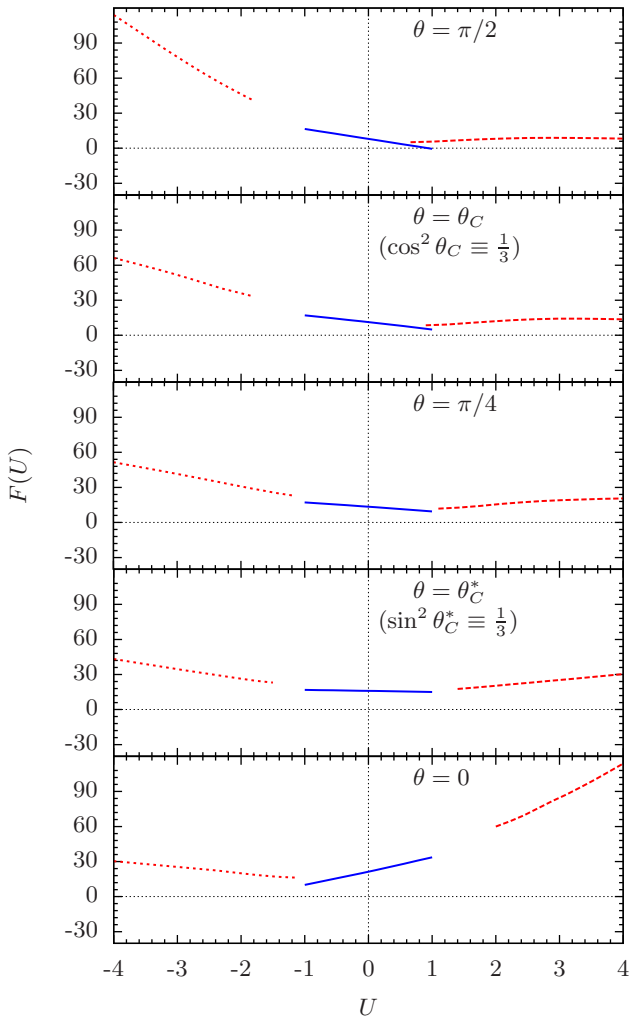}
\vspace*{0.3cm}
\caption{The function $F(U) \equiv -U^2\ln(|E|Md^2/(2\hbar^2))$ for
  different polarization angles $\theta$.  The dashed curves are
  calculated numerically and the solid lines are from
  Eq.~\eqref{e70}.}
\label{fig1}
\end{figure}

\paragraph*{Properties of weakly-bound states. }
The energy approaches zero extremely fast with vanishing strength,
where the wave function only depends on binding energy through
the modified Bessel function, $K_0(|\alpha| s)$.  The
rate of convergence towards these universal characteristics is less
clear. We therefore designed a numerical method to investigate these
structures, see \cite{vol11} for more details.

The reduced energy, $|\alpha^2|$, is about $10^{-3}-10^{-6}$ when
$U\simeq 1$, and the numerical results in Fig.~\ref{fig1} are not
easily obtained due to the exponential square dependence on $U$.  We
approach numerically the analytic straight line for small $U$, and
find a very satisfying agreement for all $\theta$. We note that the
omitted second order correction terms in Eq.~\eqref{e70} quickly, $U
\simeq 0.5$, become significant in order to reproduce the parabolic
behavior found numerically.
 
\begin{figure}[t]
\input{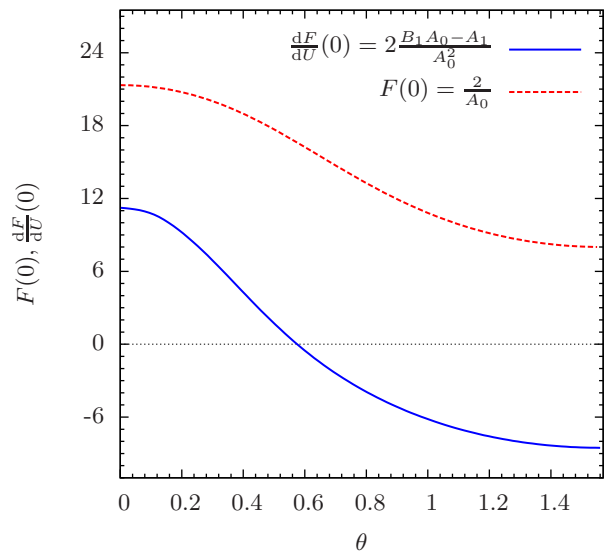}
\vspace*{0.3cm}
\caption{The value (dashed) and the derivative (solid) with respect to
  U of the function, $F(U) = -U^2\ln(|E|Md^2/(2\hbar^2))$, for $U=0$
  as functions of polarization angle.}
\label{fig2}
\end{figure}

The binding energy is from Fig.~\ref{fig1} seen to increase or
decrease with increasing $\theta$ for negative or positive $U$,
repectively. This exponential variation with $\theta$, determined by
$2/A_0$, is shown in Fig.~\ref{fig2} in the limit $U \rightarrow
0$. The approach to $U=0$ also varies strongly with $\theta$ as
illustrated in Fig.~\ref{fig2} by the slope which even changes sign at
$\theta \approx 0.57$.

The structure of the wave function is revealed through the radial
components, $f_m$, in Eq.~\eqref{e40}.  In the universal weak binding limit,
$U\rightarrow 0$, the $m=0$ solution, $K_0$, is approached for all
anisotropic potentials.  This feature of universality is reached by
smearing out the wave function over an ever increasing part of space.
Outside of the potential, there emerges a avery small and slowly
varying wave function whereas the contours of the potential are seen at
smaller distances. This variation is found even for very small $U$,
since the behavior near $r=0$ is essential to provide binding.

\begin{figure}[t]
\input{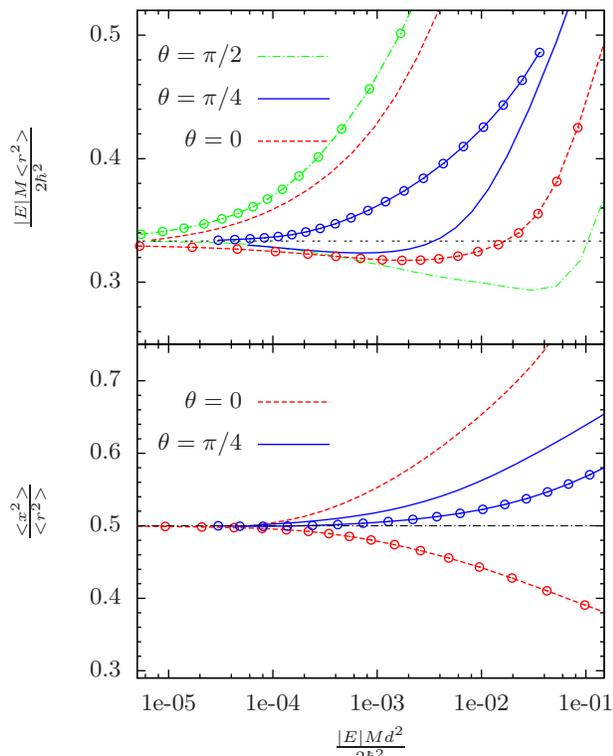}
\vspace*{0.3cm}
\caption{The upper part is energy times mean square radius, $\langle
  r^2\rangle$.  The lower part is the ratio of the second moment in
  the $x$-direction and $\langle r^2\rangle$ for different polarization
  angles $\theta$.  The energy on the $x$-axis is obtained by varying
  the strength $U$ around zero, shown for $U<0$ and $U>0$ as curves
  with and without points, respectively.  The horizontal line at $1/3$
  is the universal limit. }
\label{fig3}
\end{figure}

The convergence of the wave function towards $K_0$ implies that that
the mean square radius in the limit is inversely proportional to the
energy, $\langle r^2\rangle= 2\hbar^2/(3M|E|)$, as shown in \cite{nie01} and
illustrated in Fig.~\ref{fig3}. The limit is approached from above or
below depending on the potential.  A large barrier, (as for $U>0,
\theta= \pi/2)$, confines the wave function to small distances even
when the energy is relatively close to the threshold.  However,
eventually the wave function leaks out and the limit is approached
from below.  For $U<0, \theta \approx \pi/2$ the barrier is absent and
the approach for $E \rightarrow 0$ is from above.  Approach to the
universal limit for $E \rightarrow 0$ proves that only $m=0$ waves
remain in analogy to halo nuclei in three dimensions. For much
stronger binding $E \langle r^2\rangle$ increases as $E$ becomes large whereas
$\langle r^2\rangle$ is confined by the radius of the attractive potential.

Another feature of universality is that the deviation from cylindrical
symmetry on average disappears when $E \rightarrow 0$ as shown in
Fig.~\ref{fig3}.  The rate of disappearance varies strongly with
polarization angle.  For $\theta =\pi/2$ the potential itself is
already cylindrical, whereas the largest deformation of potential and
wave function occurs for $\theta = 0$. Increasing the angle results in
less deformation for all energies.  The elongation is largest along
the polarization direction for all $U>0$.  The deformations for $U<0$
are much smaller for comparable energies.

\paragraph*{Perspectives. }
The two-body system has to be thoroughly understood to provide a solid
ground for extraction of properties of the $N$-body system. 
We present an alternative proof that an arbitrary anisotropic 2D potential with zero net
volume has at least one bound state for infinitesimal potential
strength and rigorously derive a general expression for the energy for very weak
potentials where the system reaches universality with a wave function
given as a modified Bessel function entirely determined by the binding
energy.  The analytic
result for the energy is worked out in details for dipolar
molecules, and we illustrate the result by accurate numerical
computations.  High-order terms in the potential strength are
necessary to get accurate values.  We investigate
characteristics of universality, that is approach to cylindrical
symmetry, where the monopole component dominate and the mean square
radius becomes inversely proportional to the binding energy.  The
universal limit is reached when the reduced strength, $|U|$, decreases
below unity, which is the regime of current experiments \cite{layerexp}.

These results imply that three particles in 2D without quantum
statistical restrictions would have at least two
bound states for weak couplings with universal features \cite{nie01}.  In conclusion,
reliable predictions of many-body structures, including regions of
various phases, must take these few-body structures into account as
they persist to arbitrarily small potential strength.

\end{document}